\documentclass[11pt]{article}
\usepackage{graphicx}
\usepackage{times}
\usepackage{pifont}
\usepackage{latexsym}
\usepackage{amsfonts}
\usepackage{amsmath}
\usepackage{amssymb}

\usepackage{float}

\newcommand{\shortcite}[1]{\cite{#1}}

\newcommand{\p}{{\rm P}}
\newcommand{\np}{{\rm NP}}
\newcommand{\elec}{\electionsystem}

\newcommand{\electionsystem}{\cale}
\newcommand{\cale}{\ensuremath{\cal E}}

\newlength{\filength}
\newsavebox{\gcbox}
\sbox{\gcbox}{\framebox[\filength]{\rule{0ex}{2ex}}}

\newtheorem{theorem}{Theorem}
\newtheorem{corollary}[theorem]{Corollary}

\newcommand\qedblob{\ding{113}}
\def\literalqed{{\ \nolinebreak\hfill\mbox{\qedblob\quad}}}

\def\qed{\literalqed}

  \newtheorem{observation}[theorem]{Observation}
 \newtheorem{example}{Example}
\showhyphens{}



\newcommand{\npc}{\np\text{-complete}}

\newenvironment{proofs}{\noindent{\bf Proof.}\hspace*{1em}}{\literalqed\bigskip}
\newenvironment{proofsketch}{\noindent{\bf Proof Sketch.}\hspace*{1em}}{\literalqed\bigskip}

\DeclareMathSymbol{\subsetneq}{\mathbin}{AMSb}{"28}
\DeclareMathSymbol{\supsetneq}{\mathbin}{AMSb}{"29}

\newcommand{\score}[1]{\ensuremath{{\rm score}(#1)}}
\usepackage{enumitem}

\newcommand{\prob}[3]{
\begin{description}%
   \item[Name:] #1
   \item[Given:]  #2
   \item[Question:] #3
\end{description}}

\newcommand{\mysum}{\sum\!}

\newcommand{\Partition}{{Partition}}
\newcommand{\ExactCover}{{Exact Cover by 3-Sets}}

\begin{document}
\sloppy

\title{Complexity of Manipulative Actions When Voting with Ties}

\author{
Zack Fitzsimmons \\
  College of Computing and Inf.\ Sciences\\
  Rochester Inst.\ of Technology \\
  Rochester, NY 14623 \and
  Edith Hemaspaandra \\
  Dept.~of Computer Science\\
  Rochester Inst.\ of Technology \\
  Rochester, NY 14623}%

\date{June 15, 2015}

\maketitle

\begin{abstract}
Most of the computational study of
election problems
has assumed that each voter's preferences are,
or should be extended to, a total order.
However in practice
voters may have preferences with ties.
We study
the complexity of manipulative
actions on elections where voters can have ties,
extending the definitions of the election systems (when
necessary) to handle voters with ties.
We show that for natural election systems allowing ties 
can both increase and decrease the complexity of manipulation
and bribery, and we
state a general result on the effect of voters with ties on the complexity of
control.
\end{abstract}

\section{Introduction}
Elections are commonly used to reach a decision when presented
with the preferences of several agents. This includes political domains
as well as multiagent systems. In an election agents can have an incentive
to cast a strategic vote in order to affect the outcome.
An important negative
result from social-choice theory, the Gibbard-Satterthwaithe theorem,
states that every reasonable election system is susceptible to strategic
voting
(a.k.a.\ manipulation)~\cite{gib:j:polsci:manipulation,sat:j:polsci:manipulation}.

Although every reasonable election system can be manipulated,
it may be computationally infeasible to determine if a successful
manipulation exists.
Bartholdi et al.~\shortcite{bar-tov-tri:j:manipulating}
introduced the notion of exploring the computational complexity
of the manipulation problem. They expanded
on this work by introducing and analyzing the complexity of
control~\cite{bar-tov-tri:j:control}.
Control models the actions of an election organizer,
referred to as the chair,
who has control over the structure of the
election (e.g., the voter set) and wants to ensure
that a preferred candidate wins.
Faliszewski et al.~\shortcite{fal-hem-hem:j:bribery}
introduced the model of bribery. %
Bribery is
closely related to manipulation, but instead of asking if voters can
cast strategic votes to ensure a preferred outcome bribery asks
if a subcollection of the voters can be paid to change their vote to ensure a
preferred outcome.

It is important that we understand the complexity of these election
problems on votes that allow ties, since in practical settings voters
often have ties between some of the candidates.
This is seen in
the online preference repository \textsc{PrefLib}, which contains
several election datasets containing votes with ties, ranging from
political elections to elections created from rating
data~\cite{mat-wal:c:preflib}.
Most of the computational study of election problems for partial votes
has assumed that each voter's preferences should be extended to
a total order (see e.g., the possible and necessary winner
problems~\cite{kon-lan:c:incomplete-prefs}).
However an agent may view two options as explicitly equal and
it makes sense to view these preferences as votes with ties, instead of as partial
rankings that can be extended.

Election systems are sometimes even explicitly defined for voters
with ties. %
Both the Kemeny rule~\cite{kem:j:no-numbers}
and the Schulze rule~\cite{sch:j:clone-independent-new} are defined for
votes that contain ties. Also, there exist variants of the Borda count that
are defined for votes %
that contain ties~\cite{eme:j:partial-borda}.

The computational study of the problems of manipulation, control, and bribery
has largely been restricted to elections that contain voters with tie-free
votes.
Important recent work by
Narodytska and Walsh~\shortcite{nar-wal:c:partial-vote-manipulation}
studies the computational complexity of the manipulation problem
for top orders, i.e., votes where the candidates ranked last are all
tied and are otherwise total orders.
The manipulation results in this paper can be seen as
an extension of the work by Narodytska and Walsh.
We consider orders that allow a voter to state ties
at each position of his or her preference order,
i.e., weak orders. %
We mention that in contrast to the work by
Narodytska and Walsh~\shortcite{nar-wal:c:partial-vote-manipulation},
we give an example of a natural case where manipulation becomes
hard when given votes with ties, while it is in \p\ for total orders.
Additionally, we are the first to study the complexity of the standard models
of control and bribery for votes that contain ties. However, we mention here that
Baumeister et al.\ consider a different version of bribery called extension
bribery, for top orders
(there called top-truncated votes)~\cite{bau-fal-lan-rot:c:lazy-voters}.

The organization of this paper is as follows. In Section~\ref{sec:prelim}
we state the formal definitions and problem statements needed for our results.
The results in Section~\ref{sec:results}
are split into three sections, each showing a different behavior of voting
with ties.
In Section~\ref{sec:easy-hard} we give examples of election
systems where the problems of manipulation, bribery, and control increase in
complexity from \p\ to \npc.
Conversely, in Section~\ref{sec:hard-easy} we give
examples of election systems where the complexity of manipulation and bribery
becomes easier, and state a general result about the complexity
of control. In Section~\ref{sec:remains} we solve an open question from
Narodytska and Walsh~\shortcite{nar-wal:c:partial-vote-manipulation} and
give examples of election systems
whose manipulation complexities are unaffected by %
voters with ties.
Additionally, we completely characterize 3-candidate Copeland$^\alpha$ coalitional
weighted manipulation for rational and irrational voters with ties.
We discuss related work in Section~\ref{sec:related} and
our general 
conclusions and open directions
in Section~\ref{sec:open}.

\section{Preliminaries}\label{sec:prelim}
An \emph{election} consists of a finite set of candidates $C$ and a collection
of voters $V$, also referred to as a preference profile.
Each voter in $V$ is specified by its preference order.
We consider voters with varying amounts of ties in their
preferences. A \emph{total order} is a linear ordering of all of the candidates
from most to least preferred.
A \emph{weak order} is a transitive, reflexive, and antisymmetric ordering
where the indifference
relation (``$\sim$'') is transitive.
In general, a weak order %
can be viewed as a total order with ties.
As usual, we will colloquially refer to indifference as
ties throughout this paper since the indifference relation specifies the
preference of two elements being equal.
A \emph{top order} %
is a weak order %
with all tied candidates ranked last, and a
\emph{bottom order} is a weak order %
with all tied candidates ranked first.
In Example~\ref{ex:orders} below we present examples of each of the orders,
with and without ties, examined in this paper.
\begin{example}\label{ex:orders}
Given the candidate set $\{a, b, c, d\}$,
$a > b \sim c > d$ is a weak order, %
$a \sim b > c > d$ is a bottom order, %
$a > b > c \sim d$ is a top order, %
and $a > b > c > d$ is a total order. %
Notice that every bottom order and every top order is also a weak order, and
that every total order is also a top, bottom, and weak order.
\end{example}

An \emph{election system}, \elec, maps an election, i.e., a finite candidate set
$C$ and a collection of voters $V$, to a set of winners, where the winner set
can be any subset of the candidate set. The voters in an election can sometimes
have an associated weight where a voter with
weight $w$ %
counts as $w$ unweighted voters. %

We examine two important families of election systems,
the first being scoring rules.
A scoring rule uses a vector of the form
$\langle s_1, \ldots, s_m \rangle$, where $m$ denotes the number of
candidates, to determine each candidate's score
when given a preference profile. When the preferences are all total orders, a
candidate
at position $i$ in the preference order of a voter receives a score of $s_i$
from that voter. The candidate(s) with the highest total score win.
We consider the following three scoring rules.

\begin{description}
  \item[Plurality:] with scoring vector $\langle 1, 0, \ldots, 0 \rangle$.

  \item[Borda:] with scoring vector $\langle m-1, m-2, \ldots, 1, 0 \rangle$.
  \item[$t$-Approval:] with scoring vector $\langle \underbrace{1, \ldots, 1}_{\text{t}}, 0, \ldots, 0 \rangle$. 

\end{description}

To properly handle voters with ties in their preference orders
we define several
natural extensions which generalize the extensions from 
Baumeister et al.~\shortcite{bau-fal-lan-rot:c:lazy-voters} and
Narodytska and Walsh~\shortcite{nar-wal:c:partial-vote-manipulation}.

Write a preference order with ties as $G_1 > G_2 > \cdots > G_r$ where each
$G_i$ is a set of tied candidates.
For each set $G_i$,
let $k_i = \sum_{j=1}^{i-1} \|G_j\|$
be the number of candidates strictly preferred to every
candidate in the set. See the caption of Table~\ref{tbl:extensions} for an
example.

We now introduce the following scoring-rule extensions, which as stated above,
generalize previously used scoring-rule
extensions~\cite{bau-fal-lan-rot:c:lazy-voters,nar-wal:c:partial-vote-manipulation}.
In Table~\ref{tbl:extensions} we present an example of each of these extensions
for Borda. %
\begin{description}
  \item[Min:] Each candidate in $G_i$
  receives a score of $s_{k_i+\|G_i\|}$.
                     		  
  \item[Max:] Each candidate in $G_i$
  receives a score of $s_{k_i+1}$.
    
  \item[Round down:] Each candidate in $G_i$
      receives a score of $s_{m-r+i}$.
  \item[Average:] Each candidate in $G_i$
  receives a score of \[\frac{\sum_{j=k_i + 1}^{k_i + \|G_i\|} s_j}{\|G_i\|}.\]
                   
\end{description}

\begin{table}
\centering
\caption{The score of each candidate 
for preference order $a > b \sim c > d$
using Borda with each of our scoring-rule extensions.
We write this order as $\{a\} > \{b,c\} > \{d\}$, i.e.,
$G_1 = \{a\}$, $G_2=\{b,c\}$, and $G_3=\{d\}$.
Note that $k_1 = 0$, $k_2 = 1$, and
$k_3 = 3$.}

\begin{tabular}{ l | c | c | c | c }
Borda & $\score{a}$ & $\score{b}$ & $\score{c}$ & $\score{d}$ \\ \hline
Min  & 3 & 1 & 1 & 0 \\ 
Max & 3 & 2 & 2 & 0 \\
Round down & 2 & 1 & 1 & 0 \\
Average & 3 & 1.5 & 1.5 & 0 \\
\end{tabular}
\label{tbl:extensions}
\end{table}

The optimistic and pessimistic models from the work by Baumeister et
al.~\shortcite{bau-fal-lan-rot:c:lazy-voters} are the same as our max and min
extensions respectively, for top orders.
All of the scoring-rule extensions for
top orders found in Narodytska and
Walsh~\shortcite{nar-wal:c:partial-vote-manipulation} can be realized by
our definitions above, with our round down and average
extensions yielding
the same scores for top orders as their round down and average
extensions.
With the additional modification that $s_m = 0$
our min scoring-rule extension yields
the same scores for
top orders as round up in Narodytska and Walsh~\shortcite{nar-wal:c:partial-vote-manipulation}.

Notice that plurality using the max scoring-rule extension for bottom orders
is the same as approval voting, where each voter indicates either approval or
disapproval of each candidate and the candidate with the most approvals win.
For example, given the set of candidates $\{a,b,c,d\}$, an approval vector that
approves of $a$ and $c$, and a preference order $a \sim c > b > d$ yield the
same scores for approval and plurality using max respectively. 
In addition to scoring rules, %
elections can be defined by the pairwise majority elections between
the candidates. One important example is %
Copeland$^\alpha$~\cite{cop:m:copeland}
(where $\alpha$ is a rational number between 0 and 1), which is scored as follows.
Each candidate receives one point for each pairwise majority election
he or she wins and receives $\alpha$ points
for each tie. We also mention that Copeland$^1$ is often referred to,
and will be throughout this paper,
as Llull~\cite{hae-puk:j:electoral-writings-ramon-llull}.
We apply the definition of Copeland$^\alpha$ to weak orders in the obvious way
(as was done for top orders
in~\cite{bau-fal-lan-rot:c:lazy-voters,nar-wal:c:partial-vote-manipulation}).

We sometimes look at voters whose preferences need not be rational and we refer
to those voters as ``irrational.''
This simply means that
for every unordered pair $a,b$ of distinct candidates, the voter has $a > b$ or
$b > a$. For example, a voter's preferences could be $(a>b, b>c, c>a)$. We also
look at irrational votes with ties.

When discussing elections defined by pairwise majority elections
we sometimes refer to the \emph{induced majority graph} of a preference profile.
A preference profile $V$ where each voter has preferences over
the set of candidates $C$ induces the majority graph with a vertex set equal to
the candidate set and an edge set defined as follows.
For every $a,b \in C$ the graph contains the edge $a \to b$ if more
voters have $a > b$ than $b > a$.

\subsection{Election Problems}
We examine the complexity of the following election problems.

The coalitional manipulation problem
(where a coalition of manipulators seeks to change the outcome of the election)
for weighted voters, first
studied by
Conitzer et al.~\shortcite{con-lan-san:j:when-hard-to-manipulate},
is described below.

\prob{\elec-CWCM} %
{A candidate set $C$, a collection of
nonmanipulative voters $V$ where each voter has a positive
integral weight, a preferred candidate $p \in C$, and a collection of manipulative
voters $W$.}
{Is there a way to set the votes of the manipulators such that $p$ is an
\elec\ winner of the election $(C,V \cup W)$?}

Electoral control is the problem of determining if
it is possible for an election organizer with control over the structure of
an election,
whom we refer to as the election chair,
to ensure that a preferred candidate wins~\cite{bar-tov-tri:j:control}.
We formally define the specific control action of constructive control
by adding voters (CCAV) below. CCAV is one of the most natural cases of
electoral control and it models scenarios such as targeted voter registration
drives where voters whose votes will ensure the goal of the chair are added
to the election.

\prob{\elec-CCAV}
{A candidate set $C$, a collection of
voters $V$, a collection of unregistered voters $U$,
a preferred candidate $p \in C$, and an add limit $k \in \mathbb{N}$.}
{Is there a subcollection of the unregistered voters $U' \subseteq U$ such that
$\|U'\| \leq k$ and $p$ is an \elec\ winner of the election
$(C,V \cup U')$?}

Bribery is the problem of determining if it is possible to change the votes of
a subcollection of the voters, within a certain budget, %
to ensure that a preferred candidate wins~\cite{fal-hem-hem:j:bribery}.
The case for unweighted voters is defined below, but we also consider
the case for weighted voters.

\prob{\elec-Bribery}
{A candidate set $C$, a collection of voters $V$, a
preferred candidate $p \in C$, and a bribe limit $k \in \mathbb{N}$.}
{Is there a way to change the votes of at most $k$ of the voters in $V$
so that $p$ is an \elec\ winner?}

\subsection{Computational Complexity}\label{sec:pre-complexity}
We use the following \npc\ problems in our proofs of NP-completeness.

\prob{Exact Cover by 3-Sets}
{A nonempty set of elements $B = \{b_1, \ldots, b_{3k}\}$ and
a collection ${\cal S} = \{S_1, \ldots, S_n\}$ of 3-element subsets of $B$.}
{Does there exist a subcollection ${\cal S'}$ of ${\cal S}$ such that
every element of $B$ occurs in exactly one member of ${\cal S'}$?}   

\prob{Partition}
{A nonempty set of positive integers $k_1, \ldots, k_t$
such that $\sum_{i=1}^{t} k_i = 2K$.}
{Does there exist a subset  $A$ of $k_1, \ldots, k_t$ such that
$\mysum A = K$?\footnote{Here and elsewhere we write $\mysum A$ to denote $\sum_{a \in
A} a$.}}

Some of our results utilize %
the following variation of \Partition,
referred to as \Partition$'$,
for which we prove NP-completeness by a reduction from \Partition.

\prob{Partition$'$}
{A nonempty set of positive even integers $k_1, \ldots, k_t$ 
and a positive even integer~$\widehat{K}$.}
{Does there exist a partition $(A,B,C)$ of $k_1, \ldots, k_t$ such that
    $\mysum A = \mysum B + \widehat{K}$?}

\begin{theorem}
\Partition$'$ is $\npc$.
\end{theorem}
\begin{proofs}
The construction here is similar to the first part
of the reduction to a different version of \Partition\ from
Faliszewski et al.~\shortcite{fal-hem-hem:j:bribery}.

Given $k_1, \ldots, k_t$ such that $\sum_{i=1}^{t} k_i = 2K$, corresponding
to an instance of \Partition, we construct the following
instance $k'_1, \ldots, k'_t, \ell'_1, \ldots, \ell'_t, \widehat{K}$
of \Partition$'$.
Let $k'_i = 4^{i} + 4^{t+1} k_i$,
$\ell'_i = 4^{i}$, and
$\widehat{K} = 4^{t+1} K + \sum_{i=1}^{t} 4^{i}$.
(Note that in Faliszewski et al.~\shortcite{fal-hem-hem:j:bribery}
``3''s were used,
but we use ``4''s here so that when we add a subset of
$k'_1, \ldots, k'_t, \ell'_1, \ldots, \ell'_t, \widehat{K}$, 
we never have carries in the last $t+1$ digits base 4,
and we set the last digit to 0 to ensure that all numbers
are even.)

If there exists a partition $(A,B,C)$ of
$k'_1, \ldots, k'_t, \ell'_1, \ldots, \ell'_t$ such that
$\mysum A = \mysum B + \widehat{K}$, then $\forall i, 1 \le i \le t$,
$\lfloor (\mysum A)/4^{i} \rfloor \bmod  4 = \lfloor(\mysum B +
\widehat{K})/4^{i}\rfloor \bmod 4$.
Note that $\lfloor (\mysum A)/4^{i} \rfloor \bmod 4 = \|A \cap
\{k'_i,\ell'_i\}\|$,
$\lfloor (\mysum B)/4^{i} \rfloor \bmod 4 = \|B \cap \{k'_i,\ell'_i\}\|$,
and $\lfloor \widehat{K}/4^{i} \rfloor \bmod 4 = 1$.
So, $\|A \cap \{k'_i,\ell'_i\}\| = \|B \cap \{k'_i,\ell'_i\}\|+1$.
It follows that exactly one of $k'_i$ or $\ell'_i$ is in $A$
and neither is in $B$.
Since this is the case for every $i$,
it follows that $B = \emptyset$. Now look at all $k_i$ such that
$k'_i$ is in $A$. That set will add up to $K$, and so our
original \Partition\ instance is a positive instance.

For the converse, it is immediate that a subset $D$
of $k_1, \ldots, k_t$ that adds up to $K$ can be
converted into a solution for our \Partition$'$ instance,
namely, by putting $k'_i$ in $A$ for every $k_i$ in $D$,
putting $\ell'_i$ in $A$ for every $k_i$ not in $D$, 
letting $B = \emptyset$, and putting all other elements of
$k'_1, \ldots, k'_t, \ell'_1, \ldots, \ell'_t$ in~$C$.~\end{proofs}
\section{Results}\label{sec:results}

\subsection{Complexity Goes Up}\label{sec:easy-hard}
The
related work on the complexity of
manipulation of top orders~\cite{nar-wal:c:partial-vote-manipulation}
did not find a natural case where manipulation complexity
increases when moving from total orders to top orders.
We will show such cases in this section.

Single-peakedness is a restriction on the preferences of the voters
introduced by Black~\shortcite{bla:j:rationale-of-group-decision-making}.
Given a total order $A$ over the candidates, referred to as an axis,
a collection of voters is single-peaked with respect to $A$ if each voter has
preferences that strictly increase to a peak and
then strictly decrease, only strictly increase, or only strictly 
decrease with respect to $A$.

For our purposes we consider the model of top order single-peakedness
introduced by Lackner~\shortcite{lac:c:incomplete-sp-aaai}
where given an axis $A$, a collection of voters is single-peaked with respect to
$A$ if no voter has preferences that strictly decrease and then
strictly increase with respect to $A$.
Notice that for total orders, if a preference profile is
single-peaked with respect to Black's
model~\cite{bla:j:rationale-of-group-decision-making} it is also single-peaked
with respect to Lackner's model~\cite{lac:c:incomplete-sp-aaai}.

For single-peaked preferences we follow the model of manipulation
from Walsh~\shortcite{wal:c:uncertainty-in-preference-elicitation-aggregation}
where the axis is given and both the nonmanipulators and the manipulators all cast
votes that are single-peaked with respect to the given axis.
3-candidate Borda CWCM is known
to be in \p\ 
for single-peaked voters~\cite{fal-hem-hem-rot:j:single-peaked-preferences}. 
\begin{theorem}\cite{fal-hem-hem-rot:j:single-peaked-preferences}
3-candidate Borda CWCM for single-peaked total orders is in \p.
\end{theorem}

We now consider the complexity of 3-candidate Borda CWCM %
for top orders %
that are single-peaked. In all of our reductions the axis is $a <_A p <_A b$.
Single-peakedness with respect to this axis
allows the following top order
votes: $a > p > b$, \ $a \sim p \sim b$, \ $a > p \sim b$, \ $p > a > b$, \
$p > b > a$, \ $p > a \sim b$, \ $b > p > a$, \ and \ $b > p \sim a$.
It does not allow $a > b > p$ or $b > a > p$.
\begin{theorem}\label{thm:borda-max}
3-candidate Borda CWCM for single-peaked top orders %
using max %
is $\npc$.
\end{theorem}
\begin{proofs}
Given a nonempty set of positive
integers $k_1, \ldots, k_t$ such that $\sum_{i=1}^{t} k_i = 2K$ we
construct the following instance of manipulation.

Let the set of candidates be $C = \{a,b,p\}$. We have two nonmanipulators with the
following weights and votes.
\begin{itemize}
  \item One weight $3K$ nonmanipulator voting $a > p \sim b$.
  \item One weight $3K$ nonmanipulator voting $b > p \sim a$.
\end{itemize}
From the nonmanipulators, $\score{p} = 6K$, while $\score{a}$ and
$\score{b}$ are both $9K$.

Let there be $t$ manipulators, each with weight corresponding to an integer
from the instance of \Partition. Without loss of generality,
all of the manipulators put $p$ first.
Then $p$ receives a score of $10K$ overall. However, $a$ and $b$ can score
at most $K$ each from the votes of the manipulators, for $p$ to be
a winner.
So the manipulators must split their votes so that a subcollection of
manipulators with weight $K$ votes $p > a > b$ and a subcollection with
weight $K$ votes $p > b > a$. Notice that these are the only votes possible
to ensure that $p$ wins and that the manipulators cannot simply all vote
$p > a \sim b$ since both $a$ and $b$ receive a point from that vote
(since we are
using max) and we have no points to spare.~\end{proofs}

The above argument for max does not %
immediately apply to the other scoring-rule extensions. In particular, for min
the optimal vote for the manipulators is always to rank $p$ first and to rank
the remaining candidates tied and less preferred than $p$ %
(as in Proposition 3 of
Narodytska and Walsh~\shortcite{nar-wal:c:partial-vote-manipulation}).
So that case is in P, with an optimal manipulator vote of $p > a \sim b$.

For round down and average %
the reduction from
the proof of Theorem~\ref{thm:borda-max} will not work
since having all manipulators vote $p > a \sim b$ will make $p$ 
a winner. %
It is not hard to modify the
reduction for max %
to work for
the round-down case.
\begin{theorem}\label{thm:borda-rounddown}
3-candidate Borda CWCM for single-peaked top orders using
round-down %
is \npc.
\end{theorem}

The average scoring-rule extension case is more complicated
since it
is less close to \Partition\ than the previous
cases.  We will still be able to show NP-completeness, but we
have to reduce from the special, restricted version of \Partition\
that we defined previously in Section~\ref{sec:pre-complexity}
as \Partition$'$.\footnote{A similar
situation occurred
in the proof of Proposition 5
in Narodytska and Walsh~\shortcite{nar-wal:c:partial-vote-manipulation},
where a
(very different) specialized version of Subset Sum was constructed
to prove that 3-candidate Borda CWCM %
(in the non-single-peaked case)
for top orders using average %
remained \npc.}
\begin{theorem}\label{thm:borda-avg}
3-candidate Borda CWCM for single-peaked top orders using
average %
is \npc.
\end{theorem}
\begin{proofs}
Let $k_1, \ldots, k_t,\widehat{K}$ be an instance of \Partition$'$.
We are asking whether there exists a partition
$(A,B,C)$ of $k_1, \ldots, k_t$ such that
$\mysum A = \mysum B + \widehat{K}$.
Recall that all numbers involved are even.
Let $k_1, \ldots, k_t$ sum to $2K$.
Without loss of generality, assume that
$\widehat{K} \leq 2K$.

Let the candidates $C = \{a,b,p\}$. We have two nonmanipulators with the
following weights and votes.
\begin{itemize}
  \item One weight $6K+\widehat{K}$ nonmanipulator voting $a > p \sim b$.
  \item One weight $6K-\widehat{K}$ nonmanipulator voting $b > p \sim a$.
\end{itemize}
From the nonmanipulators, $\score{p}$ is $6K$,  $\score{a} + \score{b} =  30K$
and $\score{a} - \score{b} = 3\widehat{K}$.

Let there be $t$ manipulators, with weights $3k_1, \ldots, 3k_t$.

First suppose there exists a partition $(A,B,C)$ of 
$k_1, \ldots, k_t$ such that $\mysum A = \mysum B + \widehat{K}$.
For every $k_i \in A$, let the weight $3k_i$ manipulator
vote $p > b > a$.
For every $k_i \in B$, let the weight $3k_i$ manipulator
vote $p > a > b$.
For every $k_i \in C$, let the weight $3k_i$ manipulator
vote $p > a \sim b$.
Notice that after this manipulation that $\score{p} = 18K$,
$\score{a} = \score{b}$, and $\score{a} + \score{b} = 30K + 6K$. 
It follows that $\score{p} = \score{a} = \score{b} = 18K$.

For the converse, suppose that $p$ can be made a winner.
Without loss of generality, assume that $p$ is ranked uniquely first by all 
manipulators.
Then $\score{p} = \score{a} = \score{b} = 18K$.
Let $A'$ be
the set of manipulator weights that vote
$p > b > a$, 
let $B'$ be the set of manipulator weights
that vote $p > a > b$, and
let $C'$ be the set of manipulator weights 
that vote $p > a \sim b$.  No other votes are possible.
Let $A = \{k_i\ |\ 3k_i \in A'\}$, $B = \{k_i\ |\ 3k_i \in B'\}$,
and $C = \{k_i\ |\ 3k_i \in C'\}$.
Therefore $(A,B,C)$ corresponds to a partition of $k_1, \ldots, k_t$.
Note that $\mysum A = \mysum B + \widehat{K}$.
\end{proofs}

We now consider the complexity of CCAV, which
is one of the most natural models of control and known to
be in \p\ for plurality for total orders~\cite{bar-tov-tri:j:control}.

\begin{theorem}\label{thm:plur-ccav-total}
\cite{bar-tov-tri:j:control}
Plurality CCAV for total orders is in \p.
\end{theorem}
However below we show two cases where CCAV for plurality
is \npc\ for bottom orders and weak orders.

As mentioned in the Preliminaries, plurality using max for bottom orders
is the same as approval voting. So the theorem below immediately follows from
the proof of Theorem 4.43 from Hemaspaandra et al.~\cite{hem-hem-rot:j:destructive-control}.

\begin{theorem}\label{thm:plur-ccav-bot-max}
Plurality CCAV for bottom orders and weak orders %
using
max %
is \npc.
\end{theorem}

We now show that the case of plurality for bottom orders
and weak orders using average is \npc.

\begin{theorem}\label{thm:plur-ccav-bot-avg}
Plurality CCAV for bottom orders and weak orders %
using
average %
is \npc.
\end{theorem}
\begin{proofs}
Let $B = \{b_1, \ldots, b_{3k}\}$ and a collection
${\cal S} = \{S_1, \ldots S_n\}$ of 3-element subsets of $B$ be an instance of
\ExactCover, where
each $S_j = \{b_{j_1}, b_{j_2}, b_{j_3}\}$.
Without loss of generality let $k$ be divisible by 4 and let $\ell = 3k/4$.
We construct the following instance of control by adding voters.

Let the candidates $C = \{p\} \cup B$. Let the addition limit be $k$.
Let the collection of registered voters consist of the following $(3k^2+9k)/4$ voters.
(When ``$\cdots$'' appears at the end of a vote the remaining candidates
from $C$ are ranked lexicographically.
For example, given the candidate
set $\{a,b,c,d\}$, the vote $b > \cdots$ denotes the vote $b > a
> c > d$.) %
\begin{itemize}
  \item For each $i$, $1 \le i \le \ell$, %
	$k+3$ voters voting $b_{i} \sim b_{i+\ell} \sim b_{i+2\ell}
       \sim b_{i+3\ell} >
       \cdots$. %
  \item One voter voting $p > \cdots$. %
\end{itemize}
Let the collection of unregistered voters consist of the following $n$ voters.
\begin{itemize}
  \item For each $S_j \in {\cal S}$, one voter voting
           $p \sim b_{j_1} \sim b_{j_2} \sim b_{j_3} > \cdots$. %
\end{itemize}
Notice that from the registered voters, the score of each $b_i$ candidate
is $(k-1)/4$ greater than
the score of $p$. Thus the chair must add voters from the collection of unregistered
voters so that no $b_i$ candidate receives more than $1/4$ more points, 
while $p$ must gain $k/4$ points. Therefore the chair must add the
voters that correspond to an exact cover.
\end{proofs}

We now
present a case where the complexity of bribery goes from \p\
for total orders to \npc\ for votes with ties.
\begin{theorem}\label{thm:plur-brib-total}
\cite{fal-hem-hem:j:bribery}
Unweighted bribery for plurality for total orders is in \p.
\end{theorem}
The proof that bribery for plurality for bottom orders and weak order using max
is \npc\ immediately follows from the proof of Theorem 4.2 from
Faliszewski et al.~\shortcite{fal-hem-hem:j:bribery}, which showed bribery for
approval to be \npc.
\begin{theorem}\label{thm:plur-brib-bot-max}
Unweighted bribery for plurality for bottom orders and weak orders %
using
max %
is \npc.
\end{theorem}

\subsection{Complexity Goes Down}\label{sec:hard-easy}
Narodytska and Walsh~\shortcite{nar-wal:c:partial-vote-manipulation} show that
the complexity of coalitional manipulation can go down when moving from
total orders to top orders. %
In particular,
they show that the complexity of coalitional manipulation
(weighted or unweighted) for Borda goes from \npc\ to \p\ for
top orders using round-up. This is because in round-up an optimal manipulator
vote is to put $p$ first and have all other candidates tied for last.
In contrast,
notice that the complexity of a (standard) control action
cannot decrease when more lenient votes are allowed. This is because
the votes that create hard instances of control are still able to be cast
when more general votes are possible. The election chair is not able
to directly change votes, except in a somewhat restricted way in
candidate control cases, but it is clear to see how this does not affect the
statement below.
\begin{observation}
If a (standard) control problem is hard for a type of vote with ties, it %
remains hard for votes that allow more ties. %
\end{observation}

What about bribery?
Bribery can be viewed as a two-phase action consisting of control by deleting
voters followed by manipulation.
Hardness for a bribery problem is
typically caused by hardness of the corresponding deleting voters problem or
the corresponding
manipulation problem.  If the deleting voters problem is hard,
this problem remains hard for votes that allow ties, and it is
likely that the bribery problem
remains hard as well. Our best %
chance of finding a bribery problem that is hard for total orders
and easy for votes with ties is a problem whose manipulation problem
is hard, but whose deleting voters problem is easy.  Such problems
exist, e.g., all weighted $m$-candidate $t$-approval systems
except plurality and triviality.\footnote{By triviality we mean a scoring rule
with a scoring vector that gives each candidate the same score.}
\begin{theorem}
\cite{fal-hem-hem:j:bribery}
Weighted bribery for $m$-candidate $t$-approval for all $t \ge 2$ and $m > t$
is \npc.
\end{theorem}
For $m$-candidate $t$-approval elections (except plurality and triviality)
the corresponding
weighted manipulation problem was shown to be \npc\ by Hemaspaandra and
Hemaspaandra~\cite{hem-hem:j:dichotomy} and the corresponding deleting voters
problem was shown to be in \p\ by Faliszewski et
al.~\cite{fal-hem-hem:c:weighted-control}.

\begin{theorem}\label{thm:m-cand-t-app-wbribe}
Weighted bribery for $m$-candidate $t$-approval
for weak orders %
and for top orders using
min %
is in \p.
\end{theorem}
\begin{proofsketch}
To perform an optimal bribery, we
cannot simply perform an optimal deleting voter action
followed by an optimal manipulation action.  For example,
if the score of $b$ is already at most the score of $p$,
it does not make sense to delete a voter with vote
$b > p \sim a$.  But in the case of
bribery, we would
change this voter to $p > a \sim b$, which could
be advantageous.

However, the 
weighted constructive control by deleting voters (WCCDV)
 algorithm from~\cite{fal-hem-hem:c:weighted-control} still 
basically works.  Since $m$ is constant, there are only a constant
number of different votes possible.
And we can assume
without loss of generality that we bribe only the heaviest voters of
each vote-type and that each bribed voter is bribed to put $p$ first
and have all other candidates tied for last.   In order to find
out if there exists a successful bribery of $k$ voters,
we look at all the ways we can distribute this $k$ among the
different types of votes.  We then manipulate 
the heaviest voters of each type to put $p$ first and
have all other candidates tied for last,
and see if that makes $p$ a winner.~\end{proofsketch}

\subsection{Complexity Remains the Same}\label{sec:remains}

Narodytska and Walsh~\shortcite{nar-wal:c:partial-vote-manipulation} show that
4-candidate Copeland$^{0.5}$ CWCM
remains \npc\ for top orders.  They conjecture that this is
also the case for 3 candidates and point out that the reduction
that shows this for total orders from
Faliszewski et al.~\shortcite{fal-hem-sch:c:copeland-ties-matter}
won't work.  We will prove their conjecture,
with a reduction
similar to the proof of Theorem~\ref{thm:borda-avg}.\footnote{Menon and Larson
independently proved the top order case of the following
theorem~\cite{men-lar:t:manipulation-partial}.}
\begin{theorem}\label{thm:copeland-npc}
3-candidate Copeland$^\alpha$ CWCM remains \npc\ for top
orders, bottom orders, and weak orders, %
for all rational $\alpha \in [0,1)$ %
in the nonunique winner case (our standard model).
\end{theorem}
\begin{proofs}
Let $k_1, \ldots, k_t$ and $\widehat{K}$ be an instance of
\Partition$'$, which asks whether there exists a partition
$(A,B,C)$ of $k_1, \ldots, k_t$ such that
$\mysum A = \mysum B + \widehat{K}$.

Let $k_1, \ldots, k_t$ sum to $2K$ and without loss of
generality assume that $\widehat{K} \le 2K$.
We now construct an instance of CWCM.
Let the candidate set $C = \{a, b, p\}$ and let the
preferred candidate be $p$.
Let there be two nonmanipulators with the following
weights and votes.
\begin{itemize}
  \item One weight $K + \widehat{K}/2$ nonmanipulator voting $a > b > p$.
  \item One weight $K - \widehat{K}/2$ nonmanipulator voting $b > a > p$.
\end{itemize}
From the votes of the nonmanipulators, $\score{a} = 2$,
$\score{b} = 1$, and $\score{p} = 0$.
In the induced majority graph, there is the edge $a \to b$
with weight $\widehat{K}$, the edge $a \to p$ with weight
$2K$, and the edge $b \to p$ with weight $2K$.
Let there be $t$ manipulators with weights
$k_1, \ldots, k_t$.

Suppose that there exists a partition of $k_1,\ldots, k_t$
into $(A,B,C)$ such that $\mysum A = \mysum B + \widehat{K}$.
Then for each $k_i \in A$, have the manipulator with weight
$k_i$ vote $p > b > a$, for each $k_i \in B$, have the manipulator
with weight $k_i$ vote $p > a > b$, and for each $k_i \in C$
have the manipulator with weight $k_i$ vote $p > a \sim b$.
From the votes of the nonmanipulators and manipulators,
$\score{a} = \score{b} = \score{p} = 2\alpha$.

For the other direction, suppose that $p$ can be made a
winner. When all of the manipulators put $p$ first
then $\score{p} = 2\alpha$ (the highest score that $p$ can achieve).
Since $\alpha < 1$,
the manipulators must have %
voted such that $a$
and $b$ tie. This means that a subcollection of the manipulators with
weight $K$ voted $p > b > a$, a subcollection with weight $K-\widehat{K}$
voted $p > a > b$, and a subcollection with weight $\widehat{K}$ voted
$p > a \sim b$. No other votes would cause $b$ and $a$ to tie. %
Notice that the weights of the manipulators in the three different
subcollections form 
a partition $(A,B,C)$ of $k_1, \ldots, k_t$ such that $\mysum A = \mysum B + \widehat{K}$.
\end{proofs}

3-candidate Copeland$^\alpha$ CWCM is unusual in that the complexity can be
different if we look at the unique winner case instead of the nonunique winner
case (our standard model).
We can prove that the only 3-candidate Copeland CWCM
case that is hard for the unique winner model remains hard using a very similar
approach.

\begin{theorem}\label{thm:copeland0-npc}
3-candidate Copeland$^0$ CWCM remains \npc\ for top
orders, bottom orders, and weak orders, %
in the unique winner case. 
\end{theorem}
\begin{proofs}
Let $k_1, \ldots, k_t$ and $\widehat{K}$ be an instance of \Partition$'$, which
asks whether there exists a partition $(A,B,C)$ of $k_1, \ldots, k_t$ such that
$\mysum A = \mysum B + \widehat{K}$.

Let $k_1, \ldots k_t$ sum to $2K$ and without loss of generality assume that
$\widehat{K} \le 2K$. We now construct an instance of CWCM.
Let the candidate set $C = \{a, b, p\}$. Let the preferred candidate be $p \in
C$. Let there be two nonmanipulators with the following weights and votes.
\begin{itemize}
    \item One weight $K + \widehat{K}/2$ nonmanipulator voting $a > p > b$.
    \item One weight $K - \widehat{K}/2$ nonmanipulator voting $b > a > p$.
\end{itemize}
From the votes of the nonmanipulators $\score{a} = 2$, $\score{b} = 0$, and
$\score{p} = 1$. The induced majority graph contains the edge $a \to b$ with
weight $\widehat{K}$, the edge $a \to p$ with weight $2K$, and the edge $p \to
b$ with weight $\widehat{K}$.
Let there be $t$ manipulators with weights $k_1, \ldots, k_t$.

Suppose that there exists a partition of $k_1, \ldots, k_t$ into $(A,B,C)$ such
that $\mysum A = \mysum B + \widehat{K}$.
Then for each $k_i \in A$ have the manipulator with weight $k_i$ vote $p>b>a$,
for each $k_i \in B$ have the manipulator with weight $k_i$ vote $p > a > b$,
and for each $k_i \in C$ have the manipulator with weight $k_i$ vote
$p > a \sim b$.
From the votes of the nonmanipulators and the manipulators $\score{p} = 1$ and
$\score{a} = \score{b} = 0$.

For the other direction, suppose that $p$ can be made a unique winner. When all
of the manipulators put $p$ first then $\score{p} = 1$. So the
manipulators must have voted so that $a$ and $b$ tie, since otherwise either $a$
or $b$ would tie with $p$ and $p$ would not be a unique winner. Therefore a
subcollection of the manipulators with weight $K$ voted $p > b > a$, a
subcollection with
weight $K - \widehat{K}$ voted $p > a > b$, and a subcollection with weight
$\widehat{K}$ voted $p > a \sim b$.  No other votes would cause $a$ and $b$ to
tie.
\end{proofs}

\begin{theorem}
\label{thm:copeland-p}
3-candidate Copeland$^\alpha$ CWCM remains in $\p$ for
top orders, bottom orders, and weak orders, %
for $\alpha = 1$ for the nonunique winner case
and for all rational $\alpha \in (0,1]$ in the unique winner case. %
\end{theorem}
The proof of this theorem follows using the same arguments as the proof
of the case without ties from Faliszewski et
al.~\cite{fal-hem-sch:c:copeland-ties-matter}.
\subsubsection*{Tournament result}
We now state a general theorem on two-voter
tournaments for votes with ties.
See Brandt et al.~\shortcite{bra-har-kar-see:c:only-takes-few} for
related work on tournaments constructed from
a fixed number of voters with total orders.

\begin{theorem}\label{thm:tour-main}
A majority graph can be induced by two weak orders %
if and only if it can be induced by
two total orders.
\end{theorem}
\begin{proofsketch}
Given two weak orders $v_1$ and $v_2$ that describe preferences
over a candidate set $C$, we construct two total orders, $v'_1$ and $v'_2$
iteratively as follows.

For each pair of candidates $a,b \in C$ and $i \in \{1,2\}$, if $a > b$ in $v_i$ then
set $a > b$ in $v'_i$.

For each pair of candidates $a,b \in C$, %
if $a > b$ in $v_1$ ($v_2$) and
$a \sim b$ in $v_2$ ($v_1$) then the majority graph induced by $v_1$ and $v_2$
contains the edge $a \to b$. To ensure that the majority graph induced by $v'_1$
and $v'_2$ contains the edge $a \to b$ we must set $a > b$ in $v'_2$ ($v'_1$).

After performing the above steps there may still be a set of candidates $C'
\subseteq C$ such that $v_1$ and $v_2$ are indifferent between each pair of
candidates in $C'$.
For each pair of candidates $a,b \in C'$, $a \sim b$  in
$v_1$ and $v_2$, which implies the majority graph does not contain and edge
between $a$ and $b$. To ensure that majority graph induced by $v'_1$ and $v'_2$
does not contain an edge between $a$ and $b$, w.l.o.g.\
set $v'_1$ to strictly prefer the lexicographically smaller to the
lexicographically larger candidate and the reverse in $v'_2$.

The process described above constructs two orders $v'_1$ and $v'_2$ and
ensures that the majority graph induced by $v_1$ and
$v_2$ is the same as the majority graph induced by $v'_1$ and $v'_2$.
Since for each pair of candidates $a,b \in C$ and $i \in \{1,2\}$
we consider each possible case  where $a \sim b$ is in $v_i$ and set
either $a > b$ or $b > a$ in the corresponding order $v'_i$,
it is clear that $v'_1$ and $v'_2$ are total orders.
\end{proofsketch}

Observe that as a consequence of Theorem~\ref{thm:tour-main} we
get a transfer of NP-hardness from total orders to weak orders %
for two manipulators when the result depends
only on the induced majority graph. The proofs for
Copeland$^\alpha$ unweighted manipulation for two
manipulators for all rational $\alpha$
for total orders depend only on the induced majority
graph~\cite{fal-hem-sch:c:copeland-ties-matter,fal-hem-sch:c:copeland01},
so we can state the following corollary to Theorem~\ref{thm:tour-main}.

\begin{corollary}\label{cor:two-cope}
Copeland$^\alpha$ unweighted manipulation for two
manipulators for all rational
$\alpha \neq 0.5$ for weak orders is \npc.
\end{corollary}

\subsubsection*{Irrational voter Copeland results}
As mentioned in the preliminaries, 
another way to give more
flexibility to voters is to let the voters be irrational. %
A voter with irrational preferences can state preferences that are not
necessarily transitive and as %
mentioned in Faliszewski et al.~\cite{fal-hem-hem-rot:j:llull} a voter is likely
to posses preferences that are not transitive when making a decision based on multiple
criteria.

Additionally, the preferences of
voters can include ties as well as irrationality.
When voters are able to state
preferences that can contain irrationality and ties they can represent all possible
pairwise preferences that they may have over all of the candidates.

It is known that unweighted
Copeland$^\alpha$ manipulation is \npc\ for total orders
for all rational $\alpha$ except 0.5~\cite{fal-hem-sch:c:copeland-ties-matter,fal-hem-sch:c:copeland01}.
For irrational voters, this problem 
is in P for $\alpha = 0$, $0.5$, and $1$, and \npc\ for
all other $\alpha$~\cite{fal-hem-sch:c:copeland01}.
{\em Weighted} manipulation
for Copeland$^\alpha$ has not been studied for irrational voters.
We will do so here.

\begin{theorem}\label{thm:copeland-remains-p}
3-candidate Copeland$^\alpha$ CWCM remains in $\p$ for irrational
voters with or without ties,
for $\alpha = 1$ for the nonunique winner case
and for all rational $\alpha \in (0,1]$ in the unique winner case. %
\end{theorem}
\begin{theorem}\label{thm:copeland-remains-npc}
3-candidate Copeland$^\alpha$ CWCM remains $\npc$ for irrational voters with or
without ties, for $\alpha = 0$ in the unique winner case
and for all rational $\alpha \in [0,1)$ in the nonunique winner case.
\end{theorem}
The proofs of the above two theorems follow from the arguments in the proofs of
the corresponding rational cases, i.e., the proofs of Theorem 4.1 and 4.2
from Faliszewski et al.~\cite{fal-hem-sch:c:copeland-ties-matter} for the case
of voters without
ties and the proofs of Theorems~\ref{thm:copeland-npc},~\ref{thm:copeland0-npc}, and~\ref{thm:copeland-p}
above for the case of voters with ties.

When $\alpha = 1$, also known as Llull, interesting things happen.
It is known that 4-candidate Llull CWCM is in
P for the unique and nonunique winner
cases~\cite{fal-hem-sch:c:llull4}.
For larger fixed numbers of candidates, this is open.  Though
it is known that unweighted manipulation for Llull (with an
unbounded number of candidates) is
NP-complete in the nonunique winner case~\cite{fal-hem-sch:c:copeland01}.
In contrast, we will show now that for irrational voters, 
all these problems are in \p.

\begin{theorem}
Llull CWCM is in \p\ for irrational voters with or without ties, in the
nonunique winner case and in the unique winner case.
\end{theorem}
\begin{proofs}
Given a set of candidates $C$, a collection of voters $V$, $k$ manipulators,
and a preferred candidate $p \in C$, the preferences of the manipulators will
always contain $p > a$ for all candidates $a \neq p$. This determines the score
of $p$. In addition, let the initial preferences of the manipulators be $a > b$
for each pair of candidates $a,b \in C-\{p\}$ such that $a$ defeats $b$ in $V$
or such that $a$ ties $b$ in $V$ and $a$ is lexicographically smaller than $b$.
Note that, if $k > 0$, there are no pairwise ties in
the election with the manipulators set in this way and that the manipulators all
have strict preferences between every pair of candidates (i.e., no ties in their
preferences).
For every $a \neq p$, let ${\rm score}_{0}(a)$ be the score of $a$
with the manipulators set as above.

Construct the following flow network. The nodes are: a source $s$, a sink $t$,
and all candidates other than $p$. For every $a \in C - \{p\}$, add an edge 
with capacity ${\rm score}_{0}(a)$ from $s$ to $a$ and add an edge with capacity 
$\score{p}$ from $a$ to $t$. For every $a,b \in C-\{p\}$, add an edge from 
candidate $a$ to candidate $b$ with capacity 1 if, when all manipulators set $b >
a$, the 
score of $a$ decreases by 1 (and the score of $b$ increases by 1). 

If there is manipulation such that $p$ is a winner, then for every candidate 
$a \in C - \{p\}$, $\score{a} \le \score{p}$ so there is a network flow
that saturates all edges that go out from $s$.

If there is a network flow that saturates all edges that go out from $s$ then
for every $a,b \in C - \{p\}$ such that there is a unit of flow from $a$ to $b$,
change $a > b$ to $b > a$ in all manipulators.

This construction can be adapted to the unique winner case by letting the
capacity of the edge from $a$ to $t$ be $\score{p}-1$ instead of $\score{p}$.
\end{proofs}

\section{Related Work}\label{sec:related}

The recent work by
Narodytska and Walsh~\shortcite{nar-wal:c:partial-vote-manipulation} 
was the first paper to study the complexity of manipulation for
top orders and is very influential to our
computational study of more general votes with ties.
They studied several extensions for election systems for top orders, which we
further extend for weak orders.

Most of the related work in the computational study of election
problems assumes that the partial or tied preferences of the
voters must be extended to total orders. We mention the important work
on partial orders by Konczak and Lang~\shortcite{kon-lan:c:incomplete-prefs}
that introduces the possible and necessary winner problems. Given a preference
profile of partial votes, a possible winner is a candidate that wins in at least
one extension of the votes to total orders, while a necessary winner wins in
every extension~\cite{kon-lan:c:incomplete-prefs}.

Baumeister et al.~\shortcite{bau-fal-lan-rot:c:lazy-voters} introduced the
problem of extension bribery, 
where given voters with preferences that are top truncated,
voters are paid to extend their vote to
ensure that a preferred candidate wins.
We do not consider the problem of extension bribery, but instead
we use the standard model of bribery introduced by
Faliszewski et al.~\cite{fal-hem-hem:j:bribery}. In this model
the briber can set the entire preferences of a subcollection of voters
to ensure that a preferred candidate wins~\cite{fal-hem-hem:j:bribery}.

\section{Conclusions and Future Work}\label{sec:open}
We examined the computational complexity of the three most commonly
studied manipulative attacks on elections when voting with ties.
We found a natural case for manipulation where the complexity
increases for voters with ties, whereas it is easy for total orders.
For bribery we found examples where
the complexity increases and where it decreases.
We examined the complexity of Copeland$^\alpha$ elections with
voters with ties and even irrational votes with and without ties.
It would be interesting to see how the complexity of other
election problems are affected by voters with ties,
specifically weak orders, %
which we consider to be a
natural model for preferences in practical settings.

\section{Acknowledgments}
The authors thank Aditi Bhatt, Kimaya Kamat, Matthew Le, David Narv{\'a}ez, Amol
Patil, Ashpak Shaikh, and the anonymous referees for their helpful comments.
This work was supported in part by NSF grant no.\ CCF-1101452 and a National
Science Foundation Graduate Research Fellowship under NSF grant no.\
DGE-1102937.

\end{document}